\documentclass[amsmath,jcp,superscriptaddress,twocolumn]{revtex4-1}

\usepackage{color}
\usepackage{amsfonts}
\usepackage{amssymb}
\usepackage{graphicx}

\begin{document}
\title{Optical properties of periodically-driven open nonequilibrium quantum systems}
\author{Gabriel Cabra}
\affiliation{Department of Chemistry \& Biochemistry, University of California San Diego, La Jolla, CA 92093, USA}
\author{Ignacio Franco}
\affiliation{Department of Chemistry, University of Rochester, Rochester, NY 14627, USA}
\affiliation{Department of Physics, University of Rochester, Rochester, NY 14627, USA}
\author{Michael Galperin}
\email{migalperin@ucsd.edu}
\affiliation{Department of Chemistry \& Biochemistry, University of California San Diego, La Jolla, CA 92093, USA} 

\begin{abstract}
Characterization and control of matter by optical means is at the forefront of research both due to 
fundamental insights and technological promise. Theoretical modeling of periodically driven systems 
is a preprequisite to understanding and engineering nanoscale quantum devices for quantum technologies.
Here, we develop a theory for transport and optical response of molecular junctions, open nonequilibrium quantum systems, 
under external periodic driving. 
Periodic driving is described using the Floquet theory combined with  nonequilibrium Green's function description of the system.
Light-matter interaction is modeled employing the self-consistent Born approximation.
Generic three-level model is utilized to illustrate effect of the driving on optical and transport properties of junctions.
\end{abstract}

\maketitle

\section{Introduction}\label{intro}
Recent progress in laser techniques combined with technological advances at the nanoscale resulted in the possibility 
of optical measurements in current carrying single-molecule junctions. 
In particular, bias induced fluorescence~\cite{HoPRB08,qiu_vibrationally_2003,dong_vibrationally_2004} and 
phosphoresecence~\cite{kimura_selective_2019} were reported in the literature. Electroluminescence allowed to visualize 
intra-molecular interactions~\cite{chen_viewing_2010,zhang_visualizing_2016},
explore  real space energy transfer~\cite{imada_real-space_2016},
and probe charge fluctuations in biased molecular junctions~\cite{BerndtPRL12}.
Raman spectroscopy yields information on vibrational structure and electron flux induced heating
of vibrational and electronic degrees of freedom in single-molecule junctions~\cite{CheshnovskySelzerNatNano08,NatelsonNL08,NatelsonNatNano11,natelson_nanogap_2013}.
Recently, experimental measurements of ultra-strong light-matter interaction in single molecule cavities
(so far without current) were reported in the literature~\cite{chikkaraddy_single-molecule_2016,kongsuwan_suppressed_2018}.
Combination of molecular electronics and optical spectroscopy evolved into new direction of research coined
molecular optoelectronics~\cite{galperin_molecular_2012,galperin_photonics_2017}.

Control of response in nanoscale systems by external driving is an active area of research both experimentally and theoretically
due to technological promise of engineering and control in quantum devices. 
For example, periodic radio frequency potential modulations in a suspended carbon nanotube junction facilitated 
experimental studies of strong coupling between tunneling electron and nanomechanical motion~\cite{steele_strong_2009}.
AC-driven charging and discharging of quantum dot was employed in experimental verification of existence
of quantum stochastic resonance,
where intrinsic fluctuations lead to amplification and optimization of a weak signal~\cite{wagner_quantum_2019}.
In single molecule junctions, periodic driving by set of laser pulse pairs was suggested as a tool to study
sub-picosecond intra-molecular dynamics~\cite{selzer_transient_2013,ochoa_pump-probe_2015}.

Here we consider a molecular junction driven by a time-periodic external field. 
The effective electronic properties of this driven system are characterized by developing a theory of its optical and transport 
response in this highly nonequilibrium setting.
Theoretical studies of periodic driving in junctions usually employ quantum master 
equation~\cite{zhan_molecular_2009,donarini_a._vibration_2012,peskin_formulation_2017,al_husseini_exploring_2018,leyton_antiresonant_2018}
and nonequilibrium Green's function~\cite{park_correlation_2011,park_charge_2011,Sena_Junior_2017} approaches.
Periodicity in external driving allows mapping of the original time-dependent problem into effective
time-independent formulation in an extended Floquet space~\cite{kohler_driven_2005}.
The Floquet theory was used in numerous theoretical studies of external driving on transport  in junctions.
Studies employing combination of the theory with Schr{\" o}dinger equation~\cite{thuberg_perfect_2017},
scattering matrix approach~\cite{wu_time-dependent_2006},
quantum master equation~\cite{ho_floquet-liouville_1986,lehmann_molecular_2002,lehmann_rectification_2003,wu_noise_2010},
and Green's functions~\cite{kohler_controlling_2003,kohler_charge_2004,stefanucci_time-dependent_2008,wu_floquet-greens_2008,rai_electrically_2013}
are available in the literature.
For strongly correlated systems combinations of the Floquet with slave boson~\cite{wu_kondo_2010},
dynamical mean field theory~\cite{tsuji_correlated_2008}, and functional renormalization group~\cite{eissing_functional_2016} 
were also formulated.

Note that for molecular junctions characteristic strength of molecular coupling to contacts 
$\Gamma\sim 0.01 - 0.1$~eV\cite{kinoshita_electronic_1995} is of the same order of magnitude 
or stronger than thermal energy $k_BT\sim 0.01$~eV. Thus, Redfield/Lindblad quantum master equation,
which relies on assumption $\Gamma\ll k_BT$, is not suitable and
utilization of nonequilibrium Green's function (NEGF) technique is preferable.
We consider a case where light-matter interaction is smaller than the system-bath coupling, so that 
diagrammatic perturbation theory can be employed, and treat the interaction within the
self-consistent Born approximation (SCBA). Periodicity of driving makes the Floquet theory useful.
Here, we formulate a Floquet-NEGF-SCBA theory of transport and optical response of single-molecule junctions 
under external periodic driving.
The formulation complements study of optical  properties of current carrying junctions~\cite{galperin_optical_2006}
by accounting for periodic external driving
and extends recent study on optical absorption properties of driven matter~\cite{gu_optical_2018}
to realm of open nonequilibrium molecular systems.

The structure of the paper is the following. In Section~\ref{theory} we introduce junction model and formulate Floquet version of
the NEGF-SCBA treatment of electron and photon fluxes in the junction.
We use the Floquet-NEGF-SCBA to illustrate effects of quantum coherence on
junction responses in Section~\ref{numres}. Section~\ref{conclude} concludes and outlines directions for future research.

\begin{figure}[b]
\centering\includegraphics[width=\linewidth]{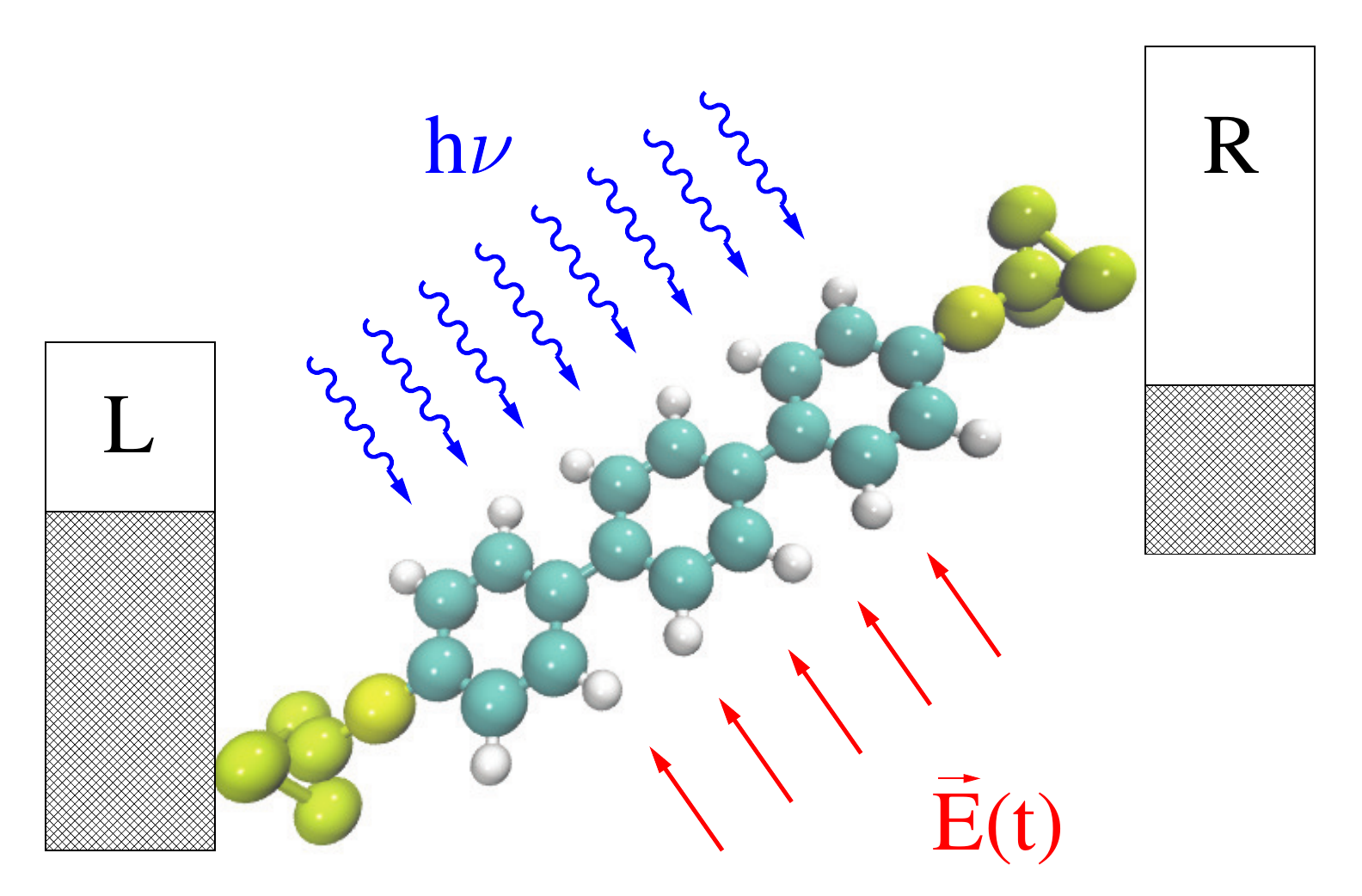}
\caption{\label{fig1}
Sketch of a molecular junction subjected to external driving.
}
\end{figure}


\section{Theory}\label{theory}
Here we present a model of a junction and briefly describe the nonequilibrium Green's function (NEGF)
approach to simulation of photon and electron fluxes, where we treat light-matter interaction  at the level of the self-consistent
Born approximation (SCBA). After this we present the Floquet formulation of the NEGF-SCBA problem.

\subsection{Model}\label{model}
We consider a junction consisting of a molecule $M$ coupled to two contacts $L$ and $R$
and to quantum modes $\{\alpha\}$ of radiation field $pt$. 
For simplicity, molecule is modeled as a non-interacting (mean-field or DFT) system.
Note, this is the usual~\cite{sergueev_ab_2005,frederiksen_inelastic_2007,avriller_inelastic_2012} 
although not always reliable~\cite{baratz_gate-induced_2013} level of modeling in first principles simulations.
In addition, the molecule is subjected to external driving by monochromatic classical field ${\vec E}(t)$
\begin{equation}
 \vec E(t)=\vec E_0\cos(\omega_0 t + \phi_0)
\end{equation}   
Contacts are modeled as reservoirs of free electrons each at its own equilibrium (see Fig.~\ref{fig1}).
The external driving distorts the electronic structure of the junction leading to effective properties 
that can in principle be very different from those observed at equilibrium.

Hamiltonian of the model is (here and below $\hbar=e=k_B=1$)
\begin{equation}
\label{Ht}
\hat H(t) = \hat H_M(t) + \sum_{B=L,R,pt}\bigg(\hat H_B + \hat V_{MB}\bigg)
\end{equation}
where $\hat H_M$ and $\hat H_B$ ($B=L,R,pt$) represent decoupled molecule and baths (contacts and radiation field),
while $\hat V_{MB}$ are the couplings. Explicit expressions are
\begin{align}
\label{Hs}
 &\hat H_M(t) = \sum_{m_1,m_2\in M} \bigg( H^M_{m_1m_2} - \vec\mu_{m_1m_2}\cdot \vec E(t) \bigg) \hat d_{m_1}^\dagger\hat d_{m_2}
 \nonumber \\
 &\hat H_{L(R)} = \sum_{k\in L(R)} \varepsilon_k\hat c_k^\dagger\hat c_k;
 \qquad
 \hat H_{pt} = \sum_{\alpha} \omega_\alpha\hat a_\alpha^\dagger\hat a_\alpha
 \\
 &\hat V_{ML(R)} = \sum_{m\in M}\sum_{k\in L(R)}\bigg( V_{km}\hat c^\dagger_k \hat d_m+ H.c.\bigg)
\nonumber \\
 &\hat V_{M,pt} = \sum_{m_1,m_2\in M} \sum_{\alpha} \bigg( 
 V^{pt}_{\alpha,m_1m_2} \hat a_\alpha^\dagger\hat d^\dagger_{m_1}\hat d_{m_2} + H.c.
 \bigg)
 \nonumber
\end{align}
Here $\hat d^\dagger_m$ ($\hat d_m$) and $\hat c_k^\dagger$ ($\hat c_k$) create (annihilate)
electron in orbital $m$ of the molecule and state $k$ of the contacts, respectively.
$\hat a_\alpha^\dagger$ ($\hat a_\alpha$) creates (destroys) excitation quanta in mode $\alpha$ of the radiation field.
$\mathbf{H}^M$ is molecular (or Kohn-Sham) orbitals basis representation of the effective single particle Hamiltonian of the molecule.
$\varepsilon_k$ and $\omega_\alpha$ are energy of free electron in state $k$ in contacts and frequency of radiation field 
mode $\alpha$, respectively. $V_{km}$ is the matrix element for electron transfer from molecular orbital $m$ to state $k$ in contacts.
$V^{pt}_{\alpha,m_1m_2}$ is the matrix element for intra-molecular (optical) electron transfer from orbital $m_2$ to orbital $m_1$
while creating excitation quanta (photon) in mode $\alpha$. 
We stress that coupling to the radiation field in (\ref{Hs}) is completely general, i.e. it is not restricted to the rotating wave approximation. 

Below for simplicity we assume 
\begin{equation}
\label{splitVpt}
V^{pt}_{m_1m_2,\alpha}=U_{m_1m_2}\, [V^{pt}_\alpha]^{*}
\end{equation}
where $\mathbf{U}$ is the matrix in molecular orbital space with $1$ for allowed optical transitions and $0$ otherwise.

\subsection{Photon and electron fluxes within NEGF-SCBA}
The nonequilibrium Green's function (NEGF) in the subspace of molecular orbitals is defined on the Keldysh contour as
\begin{equation}
 \label{defGF}
 G_{m_1m_2}(\tau_1,\tau_2)=-i\langle T_c\,\hat d_{m_1}(\tau_1)\,\hat d_{m_2}^\dagger(\tau_2)\rangle
\end{equation}
where $T_c$ is the contour ordering operator and $\tau_{1,2}$ are the contour variables.
Retarded ($r$) and lesser ($<$)/greater ($>$) projections of the Greens' function (\ref{defGF})
are usually obtained by solving differential (or Kadanoff-Baym) and integral (or Keldysh) forms of the Dyson equation, 
respectively~\cite{haug_quantum_2008,stefanucci_nonequilibrium_2013}.
\begin{align}
\label{Dyson}
&\bigg(i\,\mathbf{I}\, \frac{\partial}{\partial t_1} - \mathbf{H}^M - \vec {\mathbf{\mu}} \cdot \vec E(t_1)\bigg)\mathbf{G}^r(t_1,t_2) 
\nonumber\\ &\qquad
-\int dt\,\Sigma^r(t_1,t)\,\mathbf{G}^r(t,t_2) = \mathbf{I}\,\delta(t_1-t_2)
\\
& \mathbf{G}^{\lessgtr}(t_1,t_2) =
\nonumber \\ & \qquad
 \int dt_3\int dt_4\,\mathbf{G}^r(t_1,t_3)\,\Sigma^{\lessgtr}(t_3,t_4)\,\mathbf{G}^a(t_4,t_2)
 \nonumber
\end{align} 
Here, Hamiltonian $\mathbf{H}^M$, molecular dipole moment $\mathbf{\mu}$, 
Green's function $\mathbf{G}$ and self-energy $\Sigma$ projections are written as matrices in the subspace of molecular 
orbitals. $\mathbf{I}$ is the unity matrix in the subspace and $\mathbf{G}^a(t_4,t_2)\equiv [\mathbf{G}^r(t_2,t_4)]^\dagger$
is the advanced projection of the Green's function (\ref{defGF}).
The self-energy accounts for contributions of the baths (contacts and radiation field) into molecular dynamics
\begin{align}
\label{defSigma}
&\Sigma_{m_1m_2}(\tau_1,\tau_2) = \sum_{B=L,R,rad} \Sigma^B_{m_1m_2}(\tau_1,\tau_2) 
\nonumber \\
&\Sigma^{L(R)}_{m_1m_2}(\tau_1,\tau_2) = \sum_{k\in L(R)} V_{m_1k}\,g_k(\tau_1,\tau_2)\, V_{km_2}
\\
&\Sigma^{pt}_{m_1m_2}(\tau_1,\tau_2) = \sum_{m_3,m_4\in M} G_{m_3m_4}(\tau_1,\tau_2)
\nonumber \\ &\times
\bigg(  U_{m_3m_1}\, F(\tau_1,\tau_2)\, U_{m_4m_2}
        + U_{m_4m_2}\, F(\tau_2,\tau_1)\, U_{m_1m_3} \bigg)
\nonumber
\end{align}
where
\begin{equation}
\label{defF}
F(\tau_1,\tau_2) \equiv \sum_\alpha \lvert V^{pt}_{\alpha}\rvert^2 f_\alpha(\tau_1,\tau_2),
\end{equation} 
$g_k(\tau_1,\tau_2)\equiv -i\langle T_c\,\hat c_k(\tau_1)\,\hat c_k^\dagger(\tau_2)\rangle_0$ 
is the Green's function of free electron in state $k$, and
$f_{\alpha}(\tau_1,\tau_2)\equiv -i\langle T_c\,\hat a_\alpha(\tau_1)\,\hat a_\alpha^\dagger(\tau_2)\rangle_0$
is the Green's function of photon in mode $\alpha$ of the radiation field.

Note that while expressions for self-energies due to contacts $L$ and $R$ are exact, because of non-quadratic 
molecule-radiation field coupling in (\ref{Hs}) self-energy due to radiation field can be treated only approximately.
We utilize second order diagrammatic expansion (the self-consistent Born approximation) for $\Sigma^{pt}$
(for more details see, e.g., Ref.~\onlinecite{gao_optical_2016}).
Note also that for simplicity we disregard effect of molecule on radiation field modes.

Once Green's function is known, it can be used for evaluation of molecular responses to external perturbations. 
Below we are interested in simulating photon flux $I_{pt}(t)$~\cite{gao_optical_2016} and
electron current $I_{L\, (R)}(t)$ through interface $L\, (R)$~\cite{jauho_time-dependent_1994} 
\begin{align}
\label{fluxt}
& I_{pt}(t) = 2\,\mbox{Re}\int_{-\infty}^t dt'\,\bigg[ F^{<}(t,t')\, \Pi^{>}(t',t) 
\nonumber \\ &\qquad\qquad\qquad\qquad\;
- F^{>}(t,t')\, \Pi^{<}(t',t) \bigg]
\\
& I_{L(R)}(t) =  2\,\mbox{Re}\int_{-\infty}^t dt'\,
\mbox{Tr}\bigg[ \Sigma^{L(R)\, <}(t,t')\,\mathbf G^{>}(t',t) 
\nonumber \\ &\qquad\qquad\qquad\qquad\qquad\ \
- \Sigma^{L(R)\, >}(t,t')\,\mathbf G^{<}(t',t) \bigg]
\nonumber
\end{align}
Here, trace is over molecular subspace and $\Pi^{\lessgtr}$ is the lesser/greater 
projection of the photon self-energy due to coupling to electrons evaluated within the self-consistent Born approximation
\begin{align}
\label{defPi}
 \Pi(\tau_1,\tau_2) &= -i\, \mbox{Tr}\bigg[
 \mathbf{U}\, \mathbf{G}(\tau_1,\tau_2)\, \mathbf{U}^\dagger\,\mathbf{G}(\tau_2,\tau_1)\bigg]
\end{align}
Note that in (\ref{fluxt}) we follow the convention where positive photon flux goes from molecule to radiation field,
while positive electron flux goes from contact into the molecule.

\subsection{Floquet formulation for the NEGF-SCBA}
While for general time-dependent drivings one has to solve the Dyson equation (\ref{Dyson}) numerically on 
a two-dimensional time grid~\cite{myohanen_kadanoff-baym_2009},
periodic driving allows to formulate an effective time-independent problem in an extended Floquet space~\cite{kohler_driven_2005}. 
Because of time periodicity of the Hamiltonian,  one-sided Fourier transform of the Green's function retarded projection
can be expanded in discrete series of Floquet modes
\begin{equation}
 \label{Gr_Floquet}
 \mathbf{G}^r(t_1,t_2) = \sum_{f=-\infty}^{+\infty} \int \frac{dE}{2\pi}\, \mathbf{G}^r(f;E)\, e^{-iE(t_1-t_2)+if\omega_0 t_1}
\end{equation}
Substituting this expansion into the Dyson equations (\ref{Dyson}) leads to
\begin{align}
\label{Dyson_Floquet}
 & \bigg[\mathbf{I}(E-f\omega_0) - \mathbf{H}^M\bigg]\mathbf{G}^r(f;E)
 \nonumber \\ &
 +\frac{1}{2}\,\vec{\mathbf{\mu}}\cdot\vec E_0 \bigg[e^{i\phi_0}\mathbf{G}^r(f-1;E)+ e^{-i\phi_0}\mathbf{G}^r(f+1;E)\bigg]
\nonumber \\ &
 -\sum_{f_1,f_2=-\infty}^{+\infty}\mathbf{\Sigma}^r(f_1,f_2;E-[f+f_1]\omega_0)
 \\ &\qquad\qquad\times
 \mathbf{G}^r(f+f_1+f_2;E) 
 = \mathbf{I}\,\delta_{f,0}
\nonumber \\
 & \mathbf{G}^{\lessgtr}(f_1,f_2;E) = \sum_{f_3,f_4=-\infty}^{+\infty} \mathbf{G}^r(f_1-f_3;E-f_3\omega_0)\,
\nonumber \\ &\qquad\times
 \mathbf{\Sigma}^{\lessgtr}(f_3,f_4;E)\,\mathbf{G}^a(f_2-f_4;E-f_4\omega_0)
 \nonumber
\end{align}
Here $\mathbf{G}^a(f;E)\equiv [\mathbf{G}^r(f;E)]^\dagger$,
\begin{align}
\label{Glt_Floquet}
\mathbf{G}^{\lessgtr}(t_1,t_2) &= \sum_{f_1,f_2=-\infty}^{+\infty} \int\frac{dE}{2\pi}\,\mathbf{G}^{\lessgtr}(f_1,f_2;E)\,
\\ &\times
e^{-iE(t_1-t_2)+if_1\omega_0t_1-if_2\omega_0t_2},
\nonumber
\end{align}
and explicit expressions for the self-energy projections in the Floquet space are given in Appendix~\ref{appA}.
Equations (\ref{Dyson_Floquet}) present effective time-independent formulation of the NEGF-SCBA.

Once Floquet space projections of the Green's function are known, we use them to evaluate {\em dc}
(period averaged) photon and electron fluxes. Substituting (\ref{Gr_Floquet}) and (\ref{Glt_Floquet})
to (\ref{fluxt}) yields
\begin{align}
\label{fluxes}
 & I_{pt}^{dc} = \sum_{f=-\infty}^{+\infty} \int\frac{d\omega}{2\pi}\bigg[
 F^{<}(\omega)\,\Pi^{>}(f,f;\omega+f\omega_0) 
 \nonumber \\ &\qquad\qquad\qquad\quad\;
 - F^{>}(\omega)\,\Pi^{<}(f,f;\omega+f\omega_0) \bigg]
\nonumber \\
 & I_{L\, (R)}^{dc} = \sum_{f=-\infty}^{+\infty}\int\frac{dE}{2\pi}\,
 \\ & \qquad
 \mbox{Tr}\bigg[
 \mathbf{\Sigma}^{L(R)\, <}(E-f\omega_0)\,\mathbf{G}^{>}(f,f;E) 
 \nonumber \\ &\qquad -
 \mathbf{\Sigma}^{L(R)\, >}(E-f\omega_0)\,\mathbf{G}^{<}(f,f;E) \bigg]
 \nonumber
\end{align}
Here, $\mathbf{G}^{<}(f,f;E)$ and $\mathbf{G}^{>}(f,f;E)$ are calculated from (\ref{Dyson_Floquet}),
expressions for $\Pi^{<}(f,f;\omega)$ and $\Pi^{>}(f,f;\omega)$ are given in (\ref{Pi_Floquet}),
\begin{align}
\label{Fltgt}
& F^{<}(\omega) =  -i\,\gamma(\omega)\, N_{pt}(\omega)
\nonumber \\
& F^{>}(\omega) = -i\,\gamma(\omega)\, \left[N_{pt}(\omega)+1\right]
 \\
& \Sigma^{L(R)\, <}(E) = i\,\Gamma^{L(R)} f_{L(R)}(E)
\nonumber \\
& \Sigma^{L(R)\, >}(E) = -i\,\Gamma^{L(R)} [1-f_{L(R)}(E)] 
\nonumber
\end{align}
where 
\begin{equation}
  \label{defGamma}
  \begin{split}
& \gamma(\omega)\equiv 2\pi\sum_\alpha \lvert V^{pt}_\alpha\rvert^2\delta(\omega-\omega_\alpha)
 \\
 & \Gamma^{L(R)}_{m_1m_2}(E) \equiv 2\pi\sum_{k\in L(R)} V_{m_1 k} V_{km_2}\delta(E-\varepsilon_k)
 \end{split}
\end{equation}
are the radiative energy dissipation rate and electronic decay rate, respectively,
$f_{L(R)}(E)$ is the Fermi-Dirac distribution, and $N_{pt}(\omega)$ is population of
radiation field modes at frequency $\omega$.

We note that optical absorption theory presented in Ref.~\onlinecite{gu_optical_2018} is zero bias quasiparticle limit 
of present formulation with focus on incoming flux and neglect of self-consistency in treatment of light-matter interaction
(see Appendix~\ref{appB} for details). 
In the case of quantum treatment of radiation field,
the latter assumption is known to violate conservation laws~\cite{baym_conservation_1961,baym_self-consistent_1962}
and may lead to qualitative failures~\cite{nitzan_kinetic_2018}. 
Thus, it is not applicable in optical studies of open quantum systems~\cite{mukamel_flux-conserving_2019}. 
Below we present numerical illustrations of the developed Floquet-NEGF-SCBA formulation.

\section{Numerical results and discussion}\label{numres}
Simulations start from evaluation of Green's functions employing (\ref{Dyson_Floquet}) in the absence of radiation field
(i.e. taking self-energy $\Sigma^{pt}$ to be zero). Resulting expressions are utilized to evaluate self-energy
due to coupling to the radiation field $\Sigma^{pt}$ - last two lines in Eq.~(\ref{defSigma}). Thus obtained self-energy 
is used in simulation of the updated Green's functions employing (\ref{Dyson_Floquet}), etc.
This self-consistent procedure continues until convergence.
The solution was assumed to reach convergence when difference of level populations for each Floquet mode,
$n_m(f)\equiv-i\int dE\, G_{mm}^{<}(f,f;E)/2\pi$, at subsequent steps, $s$ and $s+1$, 
of the iterative procedure is less than $10^{-6}$: $\lvert n^{(s+1)}_m(f)-n_m^{(s)}(f)\rvert<10^{-6}$ for each $f$.

In our simulations we use a generic three level model with each level independently coupled to the contacts
\begin{equation}
\begin{split}
 H^M_{m_1m_2} &=\delta_{m_1,m_2} \varepsilon_{m_1}
 \\
 \Gamma^{L(R)}_{m_1m_2} &= \delta_{m_1,m_2} \Gamma^{L(R)}_{m_1}
 \end{split}
\end{equation}
For simplicity we also assume the wide band approximation for contacts and radiation bath, 
i.e. $\Gamma^{L(R)}$ and $\gamma$ are energy independent and corresponding Lamb shifts,
$\Lambda^{L(R)}$ and $\lambda$ are zero.
Parameters and results of the simulations are presented in terms of arbitrary unit of energy $U_0$
and corresponding unit of flux $I_0=U_0/\hbar$. 
Unless stated otherwise, parameters are the following:
$k_BT=0.03$, electron escape rates are
$\Gamma^{L}_m=\Gamma^{R}_m=0.01$ and energy dissipation rate is $\gamma=10^{-4}$.
Population of the modes corresponding to the pumping frequency $\omega_p$ is $N(\omega_p)=1$,
other modes are assumed to be unpopulated. Phase of the driving field $\phi_0=0$.
Fermi energy is taken as origin, $E_F=0$, and
bias $V_{sd}$ is applied symmetrically $\mu_{L}=E_F+|e|V_{sd}/2$ and $\mu_R=E_F-|e|V_{sd}/2$.
Floquet space in the simulations was restricted to $21$ modes, i.e. $f\in\{-10,\ldots,0,\ldots,10\}$.
Simulations are performed on energy grid spanning the range from $-2$ to $2$ with step $10^{-3}$.

\begin{figure*}[t]
\centering\includegraphics[width=\linewidth]{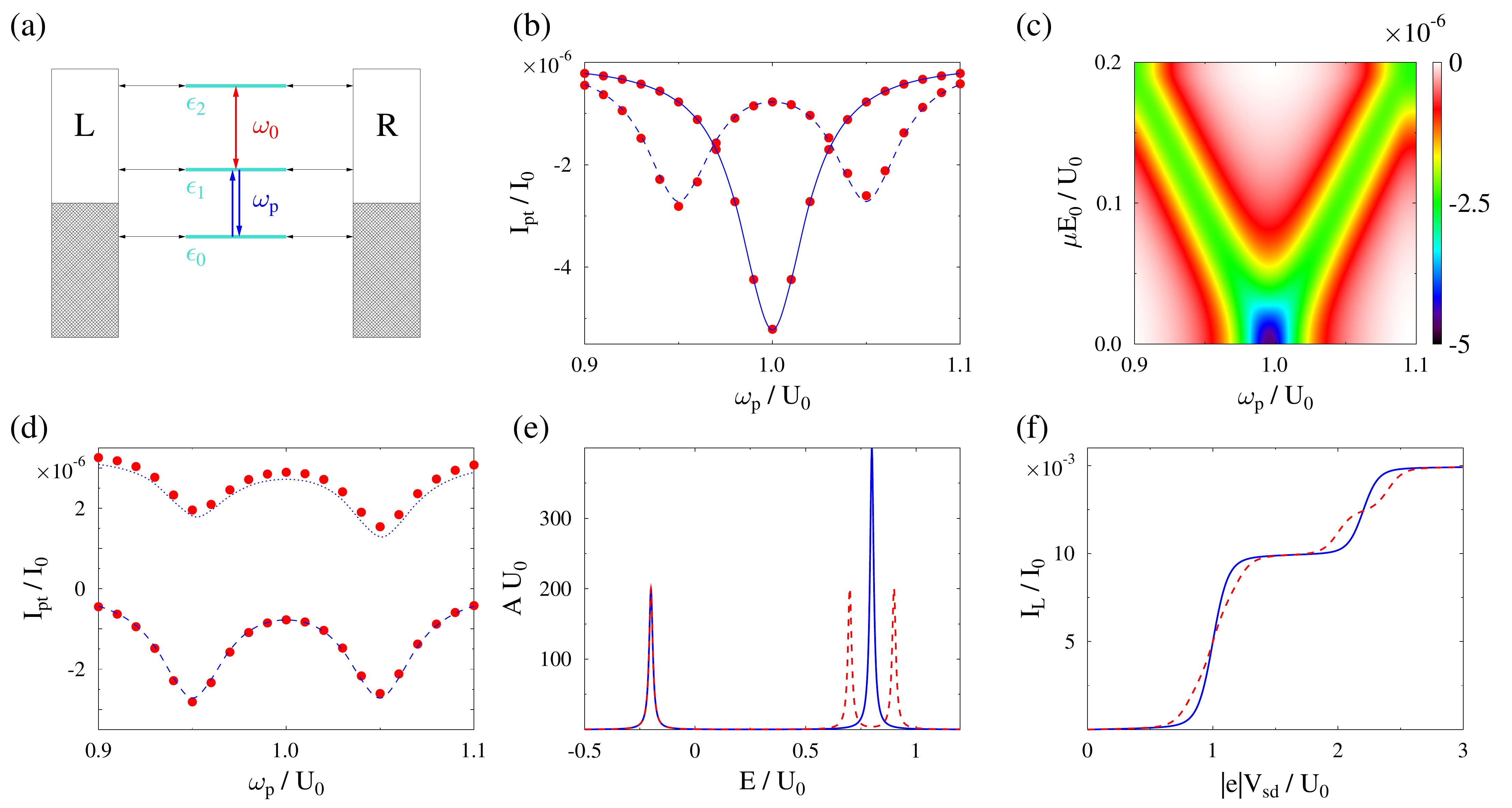}
\caption{\label{fig2}
Three-level junction responses to external drivings. 
Shown are 
(a) sketch of the junction;
(b) photon flux vs. pumping frequency  $\omega_p$ for $\vec\mu\cdot\vec E_0=0$ (solid line, blue) 
and $0.1$ (dashed line, blue) at $V_{sd}=0$;
(c) map of the photon flux vs. $\omega_p$ and $\vec\mu\cdot\vec E_0$ at $V_{sd}=0$;
(d) photon flux for $\vec\mu\cdot\vec E_0=0.1$ at $V_{sd}=0$ (dashed line, blue) and $1$ (dotted line, blue);
(e) electron density of states for $\vec\mu\cdot\vec E_0=0$ (solid line, blue) and $0.2$ (dashed line, red);
(f) electron flux vs. bias $V_{sd}$ for $\vec\mu\cdot\vec E_0=0$ (solid line, blue) and $0.2$ (dashed line, red).
Red circles in (b) and (d)  show results of the Floquet-NEGF-SCBA simulations. Other results are
obtained by going to rotational frame of the field. 
See text for parameters and further details.
}
\end{figure*}

We start from the model considered in Ref.~\onlinecite{gu_optical_2018}. 
Here, $\varepsilon_0=-0.5$, $\varepsilon_1=0.5$ and $\varepsilon_2=1.1$. 
Radiation field is assumed to be coupled to optical transition  between orbitals $0$ and $1$, 
while external driving couples orbitals $1$ and $2$ and is set to be in resonance 
with the transition, $\omega_0=0.6$ (see Fig.~\ref{fig2}a).
As discussed in Ref.~\onlinecite{gu_optical_2018},
at zero applied bias voltage ($V_{sd}=0$) single peak in absorption spectrum splits into two for stronger driving 
(compare solid and dashed lines in Fig.~\ref{fig2}b). 
The splitting starts at $\vec\mu\cdot\vec E_0\sim \sum_{m=1}^2\Gamma_m$
(here $\Gamma_m\equiv\Gamma^L_m+\Gamma^R_m$)
and increases with strength of the driving (see Fig.~\ref{fig2}c).

\begin{figure}[htbp]
\centering\includegraphics[width=\linewidth]{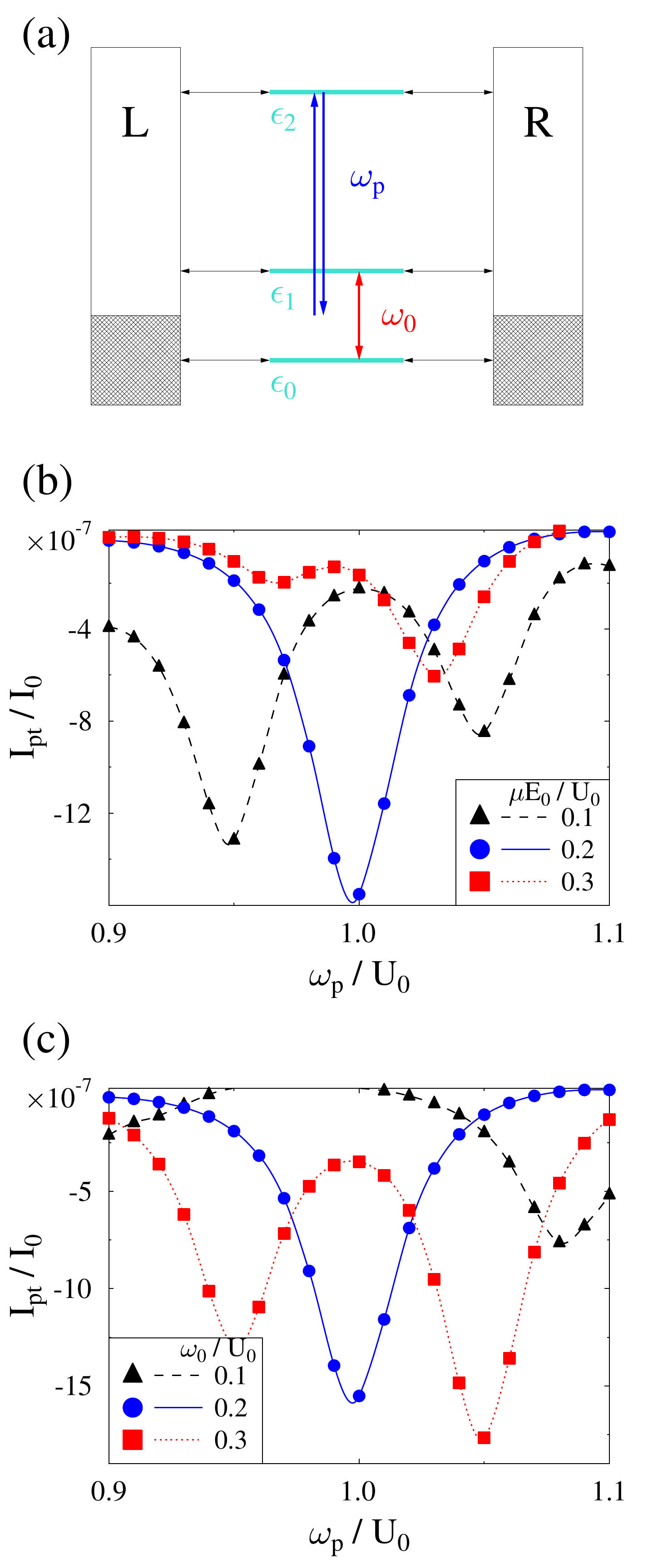}
\caption{\label{fig3}
Photon flux vs. pumping frequency  $\omega_p$ in a three-level junction model at $V_{sd}=0$. 
Shown are 
(a) sketch of the junction
(b) photon flux for several driving strengths $\vec\mu\cdot\vec E_0$ at $\omega_0=0.2$;
(c) photon flux for several driving frequencies $\omega_0$ at $\vec\mu\cdot\vec E_0=0.2$.
Simulations are performed with the Floquet-NEGF-SCBA approach. See text for parameters and further details.
}
\end{figure}

It is easy to understand the effect by going to the rotating frame of the driving field.
Because driving is at resonance, $\omega_0=\varepsilon_2-\varepsilon_1$,
utilization of the rotating wave approximation (RWA) allows to map original time-dependent problem $\hat H(t)$, 
Eq.~(\ref{Ht}), onto an effective Hamiltonian $\hat{\bar H}(t)$ 
\begin{equation}
\label{rot_frame}
\begin{split}
\hat{\bar H}(t) &= i\bigg[\frac{d}{dt} e^{\hat S(t)}\bigg] e^{-\hat S(t)} + e^{\hat S(t)} \hat H(t) e^{-\hat S(t)}  
\\
\hat S(t) &= -i\frac{\omega_0}{2}t\bigg(\hat n_0+\hat n_1-\hat n_2\bigg)
\end{split}
\end{equation}
where $\hat n_m\equiv\hat d_m^\dagger\hat d_m$.
The transformation leads to
\begin{align}
\label{barHt}
 \hat {\bar H}(t) &= \sum_{m=0}^2 \bar\varepsilon_m\hat d_m^\dagger\hat d_m
 -\frac{1}{2}\vec\mu\cdot\vec E_0\left(\hat d^\dagger_1\hat d_2 e^{i\phi_0}+\hat d_2^\dagger\hat d_1 e^{-i\phi_0}\right)
 \nonumber\\ &
 +\sum_k\varepsilon_k\hat c_k^\dagger\hat c_k + \sum_\alpha\omega_\alpha\hat a_\alpha^\dagger\hat a_\alpha
 \\ &
 +\sum_{m=0}^{2}\sum_{k}\left( \bar V_{km}(t) \hat c_k^\dagger\hat d_m + H.c.\right)
\nonumber \\ &
 +\sum_{\alpha}\left( [V^{pt}_\alpha]^{*}\hat a_\alpha^\dagger d_0^\dagger\hat d_1 + H.c.\right)
 \nonumber
\end{align}
Here $\bar\varepsilon_{0,1}=\varepsilon_{0,1}+\omega_0/2$, $\bar\varepsilon_2=\varepsilon_2-\omega_0/2$,
$\bar V_{k0(1)}(t)=V_{k0(1)}e^{-i\omega_0t/2}$ and $\bar V_{k2}(t)=V_{k2} e^{i\omega_0t/2}$.
Note that in (\ref{barHt}) the external driving enters as a time-independent electron hopping between renormalized 
degenerate levels $1$ and $2$, $\bar\varepsilon_1=\bar\varepsilon_2$.  
Moreover, because the original levels are well separated,
$\lvert\varepsilon_{m_1}-\varepsilon_{m2}\rvert\gg \Gamma_{m_1}, \Gamma_{m_2}$,
bath induced coherences among them can be safely disregarded.
In this regime, the transition to the rotating frame of the field
allows mapping onto an effective time-independent problem. Indeed, for diagonal electron dissipation matrix $\Gamma$
time factors in $\bar V_{km}(t)$ will result in shift of chemical potentials compensating renormalization of the orbital levels,
i.e. $\mu_{L(R)}$ is shifted up by $\omega_0/2$ for levels $0$ and $1$ and down by $\omega_0/2$ for level $2$. 
This time-independent problem can be treated within the standard NEGF-SCBA~\cite{galperin_inelastic_2004}.

In the rotating frame of the field, the splitting, $\vec\mu_0\cdot\vec E_0$, is manifestation of absorption into two eigenlevels resulting
from diagonalization of the $1-2$ block of the molecular Hamiltonian. The two peaks in the spectrum become distinguishable
when the splitting is bigger than broadening of the eigenlevels.
Note that for the parameters of the simulation, the RWA is quite a reasonable approximation. Effects beyond the RWA,
such as, e.g., difference absorption peak heights (dips in Fig.~\ref{fig2}b), are small - 
compare Floquet-NEGF-SCBA treatment of the original problem (circles) with NEGF-SCBA results for 
effective time independent model (dashed line).
At finite bias, when the depopulation of level $0$ and the population of level $1$ by electron transfer between molecule and contacts becomes 
possible, absorption is substituted by emission, dips become less pronounced and depths differ due to different population
of the eigenlevels at $V_{sd}=1$ (compare dashed and dotted lines in Fig.~\ref{fig2}d). 

Similarly, splitting can be observed in transport characteristics of the junction. 
Fig.~\ref{fig2}e  shows splitting in the density of states $A$ (conductance) induced by $\vec\mu\cdot\vec E_0=0.2$
external driving (compare solid and dashed lines), while Fig.~\ref{fig2}f demonstrates consequences of the splitting on 
current-voltage characteristics of the junction. The difference is due to tunneling via driven levels of the model. 
The calculations are performed in the rotated frame of the field.
 
When bath-induced coherences are non-negligible, considerations becomes more involved. In this case one cannot formulate
an effective time-independent problem, and Floquet space consideration becomes important.
For example, such situation happens when the radiation field is coupled to two optical transitions 
with external driving in between levels in the transitions.
We consider a three-level model with $\varepsilon_0=-0.1$, $\varepsilon_1=0.1$ and
$\varepsilon_2=1$. Radiation field is assumed to be coupled to optical
transitions between orbitals $0$ and $2$ and between $1$ and $2$, while external driving couples  
orbitals $0$ and $1$ (see Fig.~\ref{fig3}a). 
As previously, driving frequency is taken at resonance with its optical transition, $\omega_0=\varepsilon_1-\varepsilon_0=0.2$.
Contrary to the model in Fig.~\ref{fig2}a, where single peak splits into two on increase of driving strength,
absorption in Fig.~\ref{fig3}b show transition from split to one peak to split spectrum
(dashed to solid to dotted lines in the panel b).

Because radiation field couples orbitals $0$ and $1$ formulation of effective time-independent problem
becomes impossible. Nevertheless, transition to rotating frame similar to the one above yields
qualitative understanding of the observed effects of quantum coherence on molecular absorption spectrum.
One can show that absorption spectrum of the model in Fig.~\ref{fig3}a  having single peak (dip in Fig.~\ref{fig3})
at $\varepsilon_2-\varepsilon_0$ in the absence of driving, when the driving is present can have four peaks at frequencies
(see Appendix~\ref{appC} for details) 
\begin{equation}
\label{peaks}
\omega = \varepsilon_2-\frac{\varepsilon_0+\varepsilon_1\pm\omega_0}{2}
\pm\frac{1}{2}\sqrt{(\varepsilon_1-\varepsilon_0-\omega_0)^2+\left(\vec\mu\cdot\vec E_0\right)^2}
\end{equation}
In particular, for the parameters of calculation one expects three peaks for $\vec\mu\cdot\vec E_0=\omega_0$, and 
four peaks otherwise (two additional peaks for each $\vec\mu\cdot\vec E_0$ are outside of $\omega_p$ range in Fig.~\ref{fig3}b). 
Heights of the peaks and their broadenings can only be obtained from the simulations.
Another experimentally easily  accessible parameter is driving frequency. Also here qualitatively similar behavior
is observed (see Fig.~\ref{fig3}c).

\begin{figure}[htbp]
\centering\includegraphics[width=\linewidth]{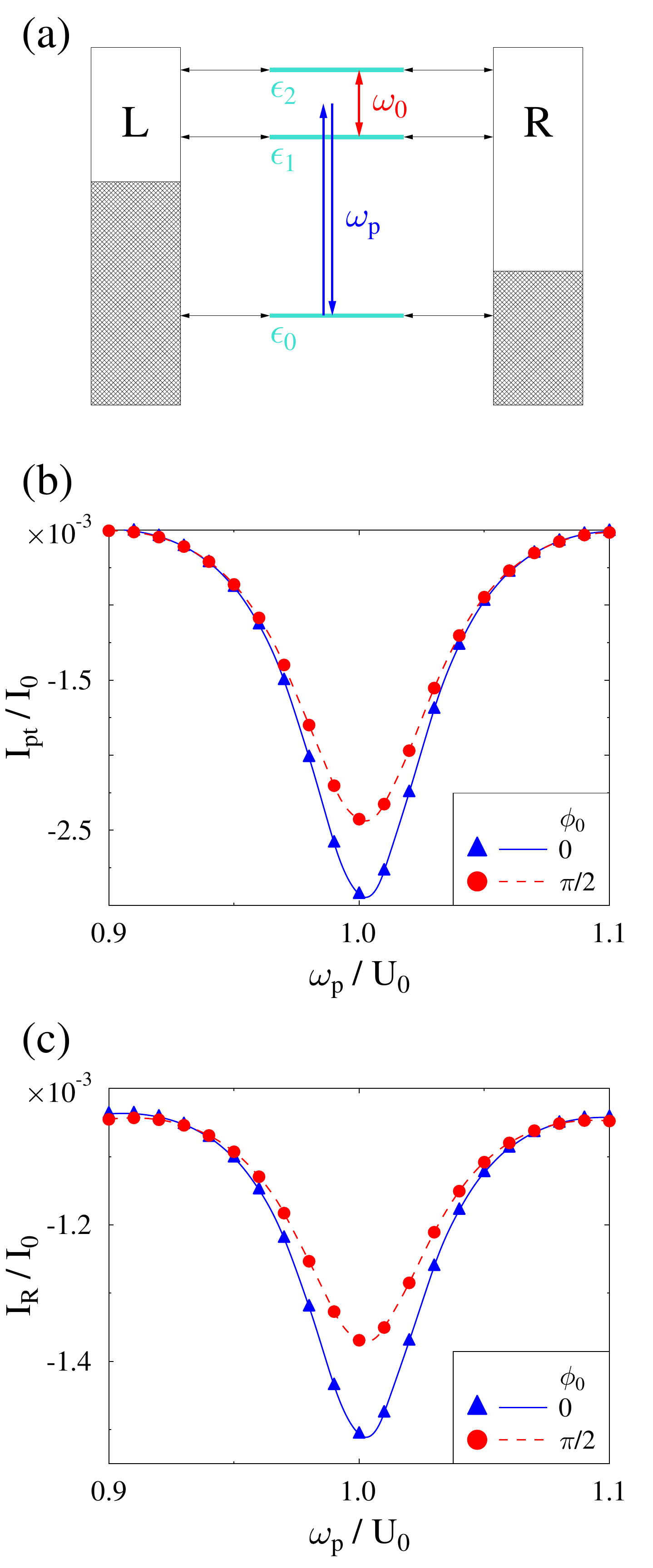}
\caption{\label{fig4}
Coherent control in a three-level junction model at $V_{sd}=0.5$. 
Shown are 
(a) sketch of the junction and (b) photon $I_{pt}$ and (c) electron $I_R$ fluxes, Eq.~(\ref{fluxes})
for $\phi_0=0$ (solid line, triangles, blue) and $\phi_0=\pi/2$ (dashed line, circles, red).
Simulations are performed with the Floquet-NEGF-SCBA approach. See text for parameters and further details.
}
\end{figure}

Finally, we demonstrate coherent control of absorption and transport in a junction beyond RWA considerations. 
The molecule is presented by its HOMO
$\varepsilon_0=-0.5$, LUMO $\varepsilon_1=0.4$ and LUMO+1 $\varepsilon_2=0.6$ levels.
External driving couples LUMO and LUMO+1 and is chosen to 
be in resonance with the optical transition, $\omega_0=0.2$. Its coupling strength is taken as 
$\vec\mu\cdot\vec E_0=0.2$. Radiation filed couples ground state (HOMO) with the two excited states 
(LUMO and LUMO+1). Energy dissipation rate is $\gamma=0.01$.
Coherence between radiation and driving fields results in control of the junction responses.
In particular, changing driving field phase $\phi_0$ one can affect optical absorption
which in turn affects current through the junction (compare solid and dashed lines in Figs.~\ref{fig4}b for absorption 
and in Fig.~\ref{fig4}c for current). Note that negative sign of photon flux indicates absorption,
which negative sign of electron flux indicates electron transfer from molecule to the contact $R$.
Note also that possibility of coherent control of optical properties was first discussed in Ref.~\onlinecite{scully_quantum_2010}.

\section{Conclusion}\label{conclude}
Characterization and control of matter by optical means is at the forefront of research both due to 
fundamental insights and technological promise. For example, in molecular junctions periodic driving of the system
by laser pulse pairs was proposed as a tool to study intra-molecular sub-picosecond dynamics in
biased junctions. Theoretical modeling of periodically driven systems  is a prerequisite to understanding 
and engineering nanoscale quantum  devices for quantum technologies.

We consider open nonequilibrium optoelectronic system under external periodic driving.
The system was treated within the nonequilibrium Green's function (NEGF) approach and 
light-matter  interaction was assumed to be small enough to allow treatment at the second order 
of the diagrammatic expansion - the self-consistent Born approximation (SCBA).
Periodic driving is accounted for within the Floquet theory, which maps original time-dependent Hamiltonian
onto effective time-independent problem in the extended Floquet space. 

Combined Floquet-NEGF-SCBA expressions were formulated and the approach was illustrated  
with generic three-level junction model simulations demonstrating coherent control of optical response in 
such devices. 
The formulation complemented study of optical  properties of current carrying junctions proposed in 
Ref~\onlinecite{galperin_optical_2006} by accounting for periodic external driving
and extended recent study on optical absorption properties of matter put forward in Ref.~\onlinecite{gu_optical_2018}
to realm of open nonequilibrium molecular systems.
Incorporation of strong intra-system interactions and application of 
the methodology in ab initio simulations are goals for future research.


\begin{acknowledgments}
This material is based upon work supported by the National Science Foundation under 
Grant No. CHE-1565939 (M.G.) and CHE-1553939 (I.F.). 
G.C. acknowledges the UCSD Physical Sciences Undergraduate Summer Research Award.
\end{acknowledgments}

\appendix
\section{Self-energies in the Floquet space}\label{appA}
Here we give expressions for the self-energies utilized in (\ref{Dyson_Floquet}).
The expressions are obtained by substituting (\ref{Gr_Floquet}) and (\ref{Glt_Floquet}) into 
corresponding projections of (\ref{defSigma}).

\begin{widetext}
Retarded projection is
\begin{align}
& \Sigma^r(f_1,f_2;E) = \delta_{f_1,0}\,\delta_{f_2,0}\, \Sigma^{L(R)\, r}(E) + \Sigma^{pt\, r}(f_1,f_2;E)
 \\
& \Sigma^{L(R)\, r}(E) = \Lambda^{L(R)}(E)-\frac{i}{2}\Gamma^{L(R)}(E)
 \\
& \Sigma^{pt\, r}(f_1,f_2;E) =  i\int \frac{d\omega}{2\pi}\bigg(
 \mathbf{U}^\dagger\bigg[\mathbf{G}^{>}(f_1,f_2;E-\omega)\, F^r(\omega)
 +\delta_{f_2,0}\mathbf{G}^r(f_1;E-\omega)\, F^{<}(\omega)\bigg]\mathbf{U}
 \\ &\qquad\qquad\qquad\qquad\qquad\;
 +\mathbf{U}\bigg[\mathbf{G}^{<}(f_1,f_2;E+\omega)\, F^a(\omega)
 +\delta_{f_2,0}\mathbf{G}^r(f_1,E+\omega)\, F^{<}(\omega)\bigg]\mathbf{U}^\dagger
 \bigg)
 \nonumber
\end{align}
\end{widetext}
where $F^{<}(\omega)$ is defined in (\ref{Fltgt}),
\begin{equation}
 F^{r}(\omega) = \lambda(\omega)-\frac{i}{2}\gamma(\omega), \quad F^a(\omega) = \left[F^r(\omega)\right]^{*},
\end{equation}
$\gamma(\omega)$ and $\Gamma^{L(R)}(E)$ are defined in (\ref{defGamma}),
and $\lambda(\omega)$ and $\Lambda^{L(R)}(E)$ are the real parts of the corresponding retarded projections
associated with $\gamma(\omega)$ and $\Gamma^{L(R)}(E)$ via the Kramers-Kronig relations.

Lesser and greater projections are
\begin{align}
 & \Sigma^\lessgtr(f_1,f_2;E) = \delta_{f_1,0}\,\delta_{f_2,0}\, \Sigma^{L(R)\, \lessgtr}(E) 
 \\ &\qquad\qquad\qquad
 + \Sigma^{pt\, \lessgtr}(f_1,f_2;E)
 \nonumber 
 \end{align}
 where $\Sigma^{L(R)\, \lessgtr}(E)$ are defined in (\ref{Fltgt}) and
 \begin{align}
 & \Sigma^{pt\, \lessgtr}(f_1,f_2;E) =  i\int \frac{d\omega}{2\pi}
\nonumber \\ & \qquad
\bigg(
 \mathbf{U}^\dagger\,\mathbf{G}^{\lessgtr}(f_1,f_2;E-\omega)\, F^{\lessgtr}(\omega)\,\mathbf{U}
 \\ & \qquad
  +\mathbf{U}\,\mathbf{G}^{\lessgtr}(f_1,f_2;E+\omega)\, F^{\gtrless}(\omega)\,\mathbf{U}^\dagger
 \bigg)
 \nonumber
\end{align}

Finally, substituting (\ref{Glt_Floquet}) into lesser and greater projections of (\ref{defPi}) leads to
\begin{align}
\label{Pi_Floquet}
 &\Pi^{\lessgtr}(f_1,f_2;\omega) = -i\sum_{\begin{subarray}{c}f_3,f_4 \\ f_5,f_6\end{subarray}=-\infty}^{_+\infty} 
 \delta_{f_1,f_3-f_5}\,\delta_{f_2,f_4-f_6}
\\ & \times
 \int\frac{dE}{2\pi}\,\mbox{Tr}\bigg[
 \mathbf{U}\,\mathbf{G}^{\lessgtr}\big(f_3,f_4;E+\frac{\omega}{2}\big)\,
 \mathbf{U}^\dagger\,\mathbf{G}^{\gtrless}\big(f_6,f_5;E-\frac{\omega}{2}\big)
 \bigg]
 \nonumber
\end{align}


\section{Comparison to the theory of Ref.~\onlinecite{gu_optical_2018}}\label{appB}
Here we compare our consideration to the theory of optical absorption presented in Ref.~\onlinecite{gu_optical_2018}.

We start from the expression for absorption rate given in Eq.~(14) of the latter reference.
In terms of our definitions the expression is
\begin{equation}
\label{appB_14}
\begin{split}
I(\omega_p) &= \lim_{t\to\infty} \frac{\lvert V^{pt}_p\rvert^2}{2\hbar^2} \frac{1}{t-t_0}\int_{t_0}^t dt_1\int_{t_0}^t dt_2
\\ &
\mbox{Re}\left[ \left(e^{-i\omega_p(t_1-t_2)}+e^{-i\omega_p(t_1+t_2)}\right) \langle \hat \mu(t_2)\,\hat\mu(t_1)\rangle \right]
\end{split}
\end{equation}
where 
\begin{equation}
\hat\mu=\sum_{m_1,m_2\in M} U_{m_1m_2} \hat d_{m_1}^\dagger \hat d_{m_2},
\end{equation}
$V^{pt}_p$ and $U_{m_1m_2}$ are defined in Eq.~(\ref{splitVpt}) and time-dependence in correlation function
of (\ref{appB_14}) is with respect to Hamiltonian excluding coupling to radiation field (interaction picture).
Note that the latter results in non-conserving expression and is applicable only in cases when 
coupling to radiation field is small (relative to other energy scales of the problem) 
so that mistake in the treatment is numerically negligible. 
Note also that factor of $2$ in the denominator in (\ref{appB_14}) comes from assumed cosine time dependence of 
the classical consideration of the radiation field in Ref.~\onlinecite{gu_optical_2018} which is effectively included in $V^{pt}_p$
in our quantum consideration. Thus, it will be dropped below. Similarly, we drop $e^{-i\omega_p(t_1+t_2)}$
because, as was discussed in Ref.~\onlinecite{gu_optical_2018}, the term does not contribute to the absorption spectrum.
Finally, we take $\hbar=1$ in our consideration. 

Implementing the adjustments and using the fact that integrand is symmetric function of $t_1$ and $t_2$,
Eq.~(\ref{appB_14}) can be rewritten as
\begin{equation}
\label{appB_14adj}
\begin{split}
I(\omega_p) &= \lim_{t\to\infty} \lvert V^{pt}_p\rvert^2 \frac{1}{t-t_0}\int_{t_0}^t dt_1\int_{t_0}^{t_1} dt_2
\\ &
2\,\mbox{Re}\left[ e^{-i\omega_p(t_1-t_2)} \langle \hat \mu(t_2)\,\hat\mu(t_1)\rangle \right]
\end{split}
\end{equation}
Below we show that this expressions yields time-averaging of in-scattering (absorption) photon flux, 
first term in Eq.~(\ref{fluxt}), i.e. we show that
\begin{equation}
\label{appB_avg}
 I(\omega_p) = -\lim_{t\to\infty} \frac{1}{t-t_0}\int_{t_0}^t dt_1 I_{pt}^{abs}(t_1)
\end{equation}
where
\begin{equation}
\label{appB_Iptabs}
 I_{pt}^{abs}(t_1) = 2\,\mbox{Re} \int_{t_0}^{t_1} dt_2 F^{<}_p(t_1,t_2)\,\Pi^{>}(t_2,t_1)
\end{equation}
Note that in our consideration $t_0$ was taken at $-\infty$ and convention of negative sign for in-coming photon flux
was followed.
Note also that for {\em dc} part of the flux (the only part considered here and in Ref.~\onlinecite{gu_optical_2018})
averaging (\ref{appB_avg}) is redundant.

Taking into account that $F^{<}$ in (\ref{fluxt}) is defined as sum over modes of radiation field, see Eq.~(\ref{defF}),
to get (\ref{appB_Iptabs}) one has to consider a single mode of the radiation field with frequency $\omega_p$.
From definition (\ref{Fltgt}) and assuming singly populated mode $N_{pt}(\omega_p)=1$,
Green's function $F^{<}_p$ for the mode can be written in time domain as
\begin{equation}
 \label{appB_Fpdef}
 F^{<}_p(t_1,t_2) = -i \lvert V^{pt}_p\rvert^2 e^{-i\omega_p(t_1-t_2)}
\end{equation}
Thus, comparing (\ref{appB_14adj}) to (\ref{appB_Iptabs}) and (\ref{appB_Fpdef}) one concludes that
\begin{equation}
\label{appB_Pidef}
 \Pi^{>}(t_2,t_1) = -i \langle \hat \mu(t_2)\,\hat\mu^\dagger(t_1)\rangle
\end{equation}
Right-hand side of this expression is definition of the two-partice Green's function.
When coupling to radiation field is neglected (as is done in Ref.~\onlinecite{gu_optical_2018}) and 
taking into account that the rest of the Hamiltonian (\ref{Hs}) does not contain many-body interactions, 
one can employ Wick's theorem expressing two-particle Green's function in terms of product of
single-particle GFs. In particular, employing Wick's theorem in the right side of (\ref{appB_Pidef})
results in an expression similar to greater projection of (\ref{defPi}). 
The crucial difference is that the resulting expression contains zero-order single-particle GFs, while Eq.~(\ref{defPi})
employs full (dressed in coupling to radiation field)  GFs.
This completes derivation of theory presented in Ref.~\onlinecite{gu_optical_2018} 
from the Floquet-NEGF-SCBA consideration. 


\section{Derivation of Eq.(\ref{peaks})}\label{appC}
Here we present qualitative analysis leading to Eq.~(\ref{peaks}).

Model shown in Fig.~\ref{fig3}a has Hamiltonian (\ref{Ht})-(\ref{Hs}) with
$H^M_{m_1m_2}=\delta_{m_1,m_2}\varepsilon_{m_1}$ ($m\in\{0,1,2\}$),
$\vec\mu_{m_1m_2}=\vec\mu(\delta_{m_1,0}\delta_{m_2,1}+\delta_{m_1,1}\delta_{m_2,0}$,
$\Gamma^{L(R)}_{m_1m_2}=\delta_{m_1,m_2}\Gamma^{(L(R)}_{m_1}$
and $U_{m_1m_2}=(\delta{m_1,0}+\delta_{m_1,1})\delta_{m_2,2}$.

Performing transformation to the field rotational frame, Eq.~(\ref{rot_frame}), with 
\begin{equation}
\hat S(t) = -i\frac{\omega_0}{2}t\bigg(\hat n_0-\hat n_1\bigg)
\end{equation}
leads to Hamiltonian similar to (\ref{barHt}) with $\bar\varepsilon_0=\varepsilon_0+\omega_0/2$,
$\bar\varepsilon_1=\varepsilon_1-\omega_0/2$ and $\bar\varepsilon_2=\varepsilon_2$;
with $\bar V_{k0}(t)=V_{k0} e^{i\omega_0t/2}$, $\bar V_{k1}(t)=V_{k1}e^{-i\omega_0t/2}$ and $\bar V_{k2}(t)=V_{k2}$;
with $\vec\mu\cdot\vec E_0$ inducing electron transfer between orbitals $0$ and $1$ 
\begin{equation}
-\frac{1}{2}\vec\mu\cdot\vec E_0 \bigg(\hat d_1^\dagger\hat d_0 e^{-i\phi_0}+ H.c.\bigg)
\end{equation}
and with coupling to radiation field - last line in Eq.~(\ref{barHt}) - becoming time-dependent
\begin{equation}
\sum_\alpha \bigg( V^{pt}_\alpha \hat d_2^\dagger\bigg[\hat d_0 e^{i\omega_0t/2}+\hat d_1 e^{-i\omega_0t/2}\bigg]\hat a_\alpha + H.c. \bigg)
\end{equation}
Diagonalizing the $0-1$ block of the molecular Hamiltonian yields egen-orbitals $\varepsilon_{+}$ and $\varepsilon_{-}$ and
leads to expressions for system-baths (contacts and radiation field)
couplings in the form
\begin{widetext}
\begin{align}
\label{appC_Vk}
& \sum_{k}\bigg(  \left[ i\, V_{k1}\sin\theta\, e^{i(\omega_0t+\phi_0)/2}- i\, V_{k0}\cos\theta\, e^{-i(\omega_0t+\phi_0)/2}\right]\hat c_k^\dagger\hat d_{+}
\\ &\quad\ \,
+\left[ i\, V_{k1}\cos\theta\, e^{i(\omega_0t+\phi_0)/2}+ i\, V_{k0}\sin\theta\, e^{-i(\omega_0t+\phi_0)/2}\right]\hat c_k^\dagger\hat d_{-}
+ V_{k_2}\hat c_k^\dagger\hat d_2 + H.c. \bigg)
\nonumber \\
\label{appC_Valpha}
& \sum_\alpha\bigg( i\, V^{pt}_\alpha \hat d_2^\dagger 
\bigg[ (\sin\theta\, e^{i(\omega_0t+\phi_0)/2}-\cos\theta\, e^{-i(\omega_0t+\phi_0)/2})\hat d_{+}
+(\cos\theta\, e^{i(\omega_0t+\phi_0)/2}+\sin\theta\, e^{-i(\omega_0t+\phi_0)/2})\hat d_{-} \bigg]
+ H.c. \bigg) 
\end{align}
\end{widetext}

Analyzing expression (\ref{appC_Vk}) within RWA-type consideration used in 
Ref.~\onlinecite{scully_degenerate_1989} and neglecting off-diagonal elements in the electronic dissipation rate matrix $\Gamma$
leads to diagonal structure of the  Green's function $G_{m_1m_2}^{(0)}=\delta_{m_1m_2} G^{(0)}_{m_1m_1}$
at zero order in the molecule-radiation field coupling.

Thus, dc part of the absorption spectrum (i.e., dc part of the incoming photon flux) takes the form
\begin{align}
& 2\,\mbox{Re}\int_{-\infty}^{t} dt'\, F^{<}(t,t')\, G^{(0)\,>}_{22}(t',t)
\\ &\times
\bigg[ 
G^{(0)\, <}_{++}(t,t')\,\bigg(\sin^2\theta\, e^{i\omega_0(t-t')/2}+\cos^2\theta\, e^{-\i\omega_0(t-t')/2}\bigg)
\nonumber \\ &\ \
+G^{(0)\, <}_{--}(t,t')\,\bigg(\cos^2\theta\, e^{i\omega_0(t-t')/2}+\sin^2\theta\, e^{-\i\omega_0(t-t')/2}\bigg)
\bigg]
\nonumber
\end{align} 
Taking integral in $t'$ leads to Eq.~(\ref{peaks})


%

\end{document}